\documentclass[12pt]{iopart}
\usepackage{epsf}
\begin{document}
\title{Quantum cryptography with correlated twin laser beams}
\author{Constantin V. Usenko\dag\ and Vladyslav C. Usenko\ddag
\footnote[3]{To whom correspondence should be addressed (cvus@ukr.net)} }

\address{\dag\ National Shevchenko University of Kyiv, Department of Theoretical Physics, Kyiv, Ukraine}

\address{\ddag\ Insitute of Physics of National Academy of Science, Kyiv, Ukraine}

\begin{abstract}
The data transmission protocol, based on the use of a strongly correlated pair of laser 
beams, is proposed.  The properties of the corresponding states are described 
in detail. The protocol is based on the strong correlation of photon numbers in both beams in each measurement.  The protocol stability against the interception attempts is 
analyzed.
\end{abstract}
%\submitto{\JOB}

\section{Introduction.}

The main goal of the quantum cryptography, which is the part of the quantum 
computing, is in development of the reliable and secure 
procedures of generation and transmission of a cryptographic key, which can 
be used for the encryption of the further communication. 
In the last years, some progress was 
achieved in this area \cite{qc1, qc2, qc3} and protocols were 
developed on the basis of the quantum entanglement \cite{ent1, entprot} of weak beams and the 
sigle \cite{single1, single2, Grangier3} or four photon states \cite{four}, mostly by means of adjusting and detecting their polarization angles \cite{polar}. 

Those methods were realized experimentally \cite{entprot, fourexp}, but still they are difficult for implementation in particular because of the complexity of a few photon 
state preparation and detection. If a bit is transferred by a one or a few 
photons, the detection of each state requires numerous acts of measurements, this
slows down the information flow. This relates even to the most successful realizations 
\cite{Grangier1, Grangier2} and that's why the new ideas are still in need. 

% pravka 1 (vyshe)

In this work we propose and examine the cryptographic method based on the 
use of a correlated two-mode laser beam for a secure key generation and transmission 
between two sites. Such and similar beams are actively experimentally 
studied last time \cite{similar1, similar2, similar3}. Therefore we examine in detail the properties 
of the states, which describe the two-mode correlated laser beams, investigate the 
dependence of these properties on the beam intensity, and analyze the possibility to use such beams in the data channels. Also we study the question of stability of 
such channels against the elementary eavesdropping attacks.

\section{The coherently correlated state}

The two-mode coherently correlated state is the way we refer to the 
generalized coherent state in the meaning by Perelomov \cite{per}. 
Such states were studied by Arvind \cite{two1} and others \cite{two2,two3,two4} as the pair-coherent states.

%pravka 2 (vyshe)

The two-mode coherently correlated state can be described by its presentation through series by Fock states:

\begin{equation}
\label{eq:tmcc}
\left| \lambda \right\rangle =\frac{1}{\sqrt{I_0\left(2\left|\lambda\right|\right)}}\sum {\frac{\lambda ^n}{n!}\left| {nn} 
\right\rangle } 
\end{equation}

Here we use the designation $\left| {nn} \right\rangle = \left| n 
\right\rangle _1 \otimes \left| n \right\rangle _2 $, where $\left| n 
\right\rangle _1 $ and $\left| n \right\rangle _2 $stand for the states of 
the $1^{st}$ and $2^{nd}$ modes accordingly, represented by their photon 
numbers. The states (\ref{eq:tmcc}) are not the eigenstates for each of the operators 
separately, but are the eigenstates for the product of annihilation 
operators: 
\begin{equation}
	a_1 a_2 \left| \lambda \right\rangle = \lambda \left| \lambda 
	\right\rangle . 
\end{equation}
Such states can also be obtained from the zero state:

\begin{equation}
\label{eq:tmccground}
\left| \lambda \right\rangle = \frac{1}{\sqrt{I_0\left(2\left|\lambda\right|\right)}}I_0 (\lambda a_1^ + a_2^ + )\left| 0 
\right\rangle 
\end{equation}

Hereinafter we denote the two-mode coherently correlated states as the 
\emph{TMCC} states. In this work we assume that two laser beams, which are 
propagating independently from each other, correspond to the two modes of 
the TMCC state. States of beams are mutually correlated. (Surely, the TMCC 
state can also be represented in another way, for example, as a beam 
consisting of two correlated polarizations)

An observable of such a pair of beams (for example, the vector-potential) is 
given by the expression:

\begin{equation}
\label{eq:vectpot}
A = \varphi _1^\ast (x,t)a_1^ + + \varphi _1 (x,t)a_1 + \varphi _2^\ast 
(x,t)a_2^ + + \varphi _2 (x,t)a_2 
\end{equation}

This expression has explicit spatial dependence $\varphi (x,t)$ and the 
quantum operators $a,a^+ $.

Let's compare the TMCC state to the usual, noncorrelated two-mode coherent 
state $| \alpha \rangle = |  \alpha_1 \rangle _1 \otimes | \alpha_2 \rangle _2$. 
Each of the two modes of such state is given by an expression:

\begin{equation}
\label{eq:usual}
\left| {\alpha _i } \right\rangle = e^{ - \frac{\left|\alpha\right|^2}{2}}\sum 
{\frac{\alpha _i^n }{\sqrt {n!} }\left| n \right\rangle } , i = 1,2
\end{equation}

Such states are the eigenstates for the corresponding annihilation 
operators: $a_i \left| {\alpha _i } \right\rangle = \alpha _i \left| {\alpha 
_i } \right\rangle $

Thus the mean value of the vector-potential (\ref{eq:vectpot}) is

\begin{equation}
\left\langle A \right\rangle = \varphi _1 \alpha _1 + \varphi _1 ^\ast 
\alpha _1 ^\ast + \varphi _2 \alpha _2 + \varphi _2 ^\ast \alpha _2 ^\ast 
\end{equation}

\noindent
and this is the show of the quasiclassical properties of the beam (\ref{eq:usual}).

In the case of the TMCC state the mean value of any characteristic, which is 
linear in field, turns to be equal to 0, because during the averaging by the 
1$^{st}$ mode the $a_1$ converts $\left| {n,n} \right\rangle $ to $\left| 
{n - 1,n} \right\rangle $, which is orthogonal to all the present state 
terms, so $\left\langle {\lambda _i } \right|a_i \left| {\lambda _i } 
\right\rangle = 0$, that's why

\begin{equation}
\left\langle A \right\rangle =\left\langle \lambda \right|A\left| \lambda 
\right\rangle =\left\{
\begin{array}{l}
  \varphi _1^\ast (x,t)\left\langle \lambda \right|a_1^ + 
\left| \lambda \right\rangle + \varphi _1 (x,t)\left\langle \lambda 
\right|a_1 \left| \lambda \right\rangle + \\
 \varphi _2^\ast (x,t)\left\langle 
\lambda \right|a_2^ + \left| \lambda \right\rangle + \varphi _2 
(x,t)\left\langle \lambda \right|a_2 \left| \lambda \right\rangle 
\end{array}\right.
= 0
\end{equation}

\noindent
and so the quasiclassical properties in their usual meaning are absent in 
this case. But they become apparent in the spatial correlation function

\begin{equation}
\gamma (x,t;{x}',{t}') = \left\{
\begin{array}{l}
 \left\langle {A(x,t) \cdot A({x}',{t}')} 
\right\rangle - \left\langle {A(x,t)} \right\rangle \cdot \left\langle 
{A({x}',{t}')} \right\rangle \\
= \left\langle {A(x,t) \cdot A({x}',{t}')} \right\rangle 
\end{array} \right. ,
\end{equation}

\noindent
which is non-zero because $\left\langle {A(x,t) \cdot A({x}',{t}')} 
\right\rangle $ contains mean values for the products of quantum operators 
and some of them are non-zero.

\begin{equation}
\label{corrav}
\left\langle {
A\left(x,t\right) \cdot A\left(x',t'\right)}
\right\rangle ={\left\{
\begin{array}{l}
 \varphi_1^\ast \left(x,t\right)\varphi_1 \left(x',t'\right) 
\left\langle \lambda \right| \hat{a}_1^{+} \hat{a}_1 \left| \lambda \right\rangle +\\
 \varphi _1^\ast \left(x,t\right)\varphi _2^\ast\left(x',t'\right)
\left\langle \lambda \right|\hat{a}_1^{+} \hat{a}_2^{+} \left| \lambda \right\rangle+\\
 \varphi _1 \left(x,t\right)\varphi _1^\ast \left(x',t'\right)
\left\langle \lambda \right|\hat{a}_1 \hat{a}_1^{+} \left| \lambda \right\rangle+ \\ 
 \varphi_1\left(x,t\right)\varphi_2\left(x',t'\right)
\left\langle \lambda \right|\hat{a}_1\hat{a}_2 \left| \lambda \right\rangle+ \\
\varphi _2^\ast \left(x,t\right)\varphi _1^\ast \left(x',t'\right)
\left\langle \lambda \right|\hat{a}_2^{+} \hat{a}_1^{+} \left| \lambda \right\rangle+\\
 \varphi _2^\ast \left(x,t\right)\varphi _2 \left(x',t'\right)\left\langle 
\lambda \right|\hat{a}_2^{+} \hat{a}_2 \left| \lambda \right\rangle +\\
 \varphi_2 \left(x,t\right)\varphi_1 \left(x',t'\right)
\left\langle \lambda \right|\hat{a}_2 \hat{a}_1 \left| \lambda \right\rangle+\\ 
 \varphi _2 \left(x,t\right)\varphi _2^\ast \left(x',t'\right)
 \left\langle \lambda \right|\hat{a}_2 \hat{a}_2^{+} \left| \lambda \right\rangle 
\end{array}\right.}
\end{equation}

\section{Communication via quantum channel}

\begin{figure}[htbp]
	\epsfbox{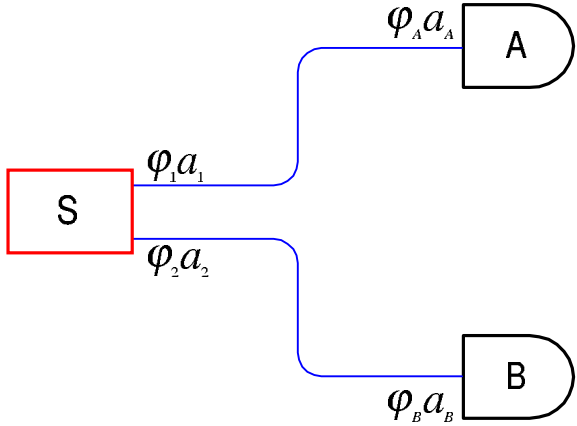}
	\caption{Quantum channel between two parties with a TMCC source}
	\label{fig1}
\end{figure}

Let we have to establish a secure quantum channel between two parties (\Fref{fig1}). Alice has the laser on her side, which produces two beams in the TMCC 
state. The optical channel is organized in such a way, that Alice receives one 
of the modes, the first, for example, i.e. $\varphi _A \equiv \varphi _1 
$,$\varphi _A (x_A ,t_0 ) = 1$ , and Bob receives another one, i.e. $\varphi _B 
\equiv \varphi _2 $ ,$\varphi _B (x_B ,t_0 ) = 1$ at any moment of 
measurement $t_0 $, where $x_A $and $x_B $are Alice's and Bob's locations 
respectively. Accordingly, Alice cannot measure the Bob's beam and vice 
versa:$\varphi _B (x_A ,t_0 ) = 0$, $\varphi _A (x_B ,t_0 ) = 0$. At that 
the field is:

\begin{equation}
A = \varphi _A^\ast (x,t)a_A^ + + \varphi _A (x,t)a_A + \varphi _B^\ast 
(x,t)a_B^ + + \varphi _B (x,t)a_B 
\end{equation}

The intensity of the radiation, registered by Alice is proportional to the 
mean of the $N_A = a_A^ + a_A$ operator, which is the number of the photons 
in the $1^{st}$ mode and it is similarly for Bob with $N_B = a_B^ + a_B$ . 
Thus the mean observable values, which characterize the results of the 
measurements, taken by Alice and Bob, are

\begin{equation}
\left\langle {N_{A,B} } \right\rangle = \left\langle \lambda \right|a_{A,B}^ 
+ a_{A,B} \left| \lambda \right\rangle = \frac{\left| \lambda^2 \right|I_1 
\left( {2\left| \lambda \right|} \right)}{I_0 \left( {2\left| \lambda 
\right|} \right)}
\end{equation}

These values are squared in field, and thus their mean values don't turn to 
zero.

\begin{equation}
\label{eq:meansqrn}
\left\langle {N_{A,B}^2 } \right\rangle = \left\langle {N_A N_B } 
\right\rangle = \left\langle \lambda \right|a_A^ + a_A a_B^ + a_B \left| 
\lambda \right\rangle = \left| \lambda \right|^2
\end{equation}

The measurements have the statistical uncertainty, caused by quantum 
fluctuations. For each of the observers the uncertainty can be characterized 
by the corresponding dispersion:

\begin{equation}
\sigma _{A,B}^2 = \left\langle {N_{A,B}^2 } \right\rangle - \left\langle 
{N_{A,B} } \right\rangle ^2
\end{equation}

Taking into account (\ref{eq:meansqrn}), we get the following 
expression:

\begin{equation}
\label{eq:sigma}
\sigma _{A,B}^2 = \left| \lambda \right|^2\left( {1 - \left( {\frac{I_1 
\left( {2\left| \lambda \right|} \right)}{I_0 \left( {2\left| \lambda 
\right|} \right)}} \right)^2} \right)
\end{equation}

The interdependence of the results of measurements taken by Alice and Bob 
can by characterized by the correlation function:

\begin{equation}
g_{AB} = < N_A N_B > - < N_A > < N_B > 
\end{equation}

It's useful to describe the channel quality by the relative correlation, 
which is

\begin{equation}
\rho _{AB} = \frac{ < N_A N_B > - < N_A > < N_B > }{\sigma _A \sigma _B }
\end{equation}

The main feature of the TMCC state is that the value $\rho _{AB} $ is 
exactly equal to 1, while in the case of non-correlated beams we would get 
$\rho _{AB} = 0$. This means that the measurements of the photon numbers, 
got by Alice and Bob, each with her/his own detector, not only show the same 
mean values, but even have the same deflection from the mean values.

The laser beam is the semi-classical radiation with well defined phase, but 
due to the uncertainty principle for the number of photons and the 
phase of the radiation, there is a large enough uncertainty in the photon 
numbers, this can be seen from the dispersion expression (\ref{eq:sigma}). Thus one can 
observe the noise, which is similar to the shot noise in an electron tube. 
In the TMCC radiation the characteristics of such noise for each of the 
modes are amazingly well correlated to each other. This fact 
enables the use of such radiation for generation of a random code, which 
will be equally good received by two mutually remote detectors.

\section{The protocol}

We propose the following scheme for the TMCC-based protocol. The laser is 
set up to produce the constant mean number of photons during the session and 
both parties know this number. At some moment Alice and Bob start the 
measurements. They detect the number of photons at unit time  by measuring 
the integrated intensity of the corresponding incoming beam. If the number 
of photons for the specific unit time is larger than the known expected mean 
(which is due to the shot noise), the next bit of the generated code is 
considered to have the value ``1''. If the measured number is less than 
the expected mean, the next bit is considered to be equal to ``0''. The 
procedure is repeated until both, Alice and Bob, get enough bits for the 
cryptographic key. The described protocol can be supplied with the 
procedures of the cryptographic control.

\section{Eavesdropping}

Since the proposed protocol uses the scheme, which differs from the 
well-known schemes, based on the entangled states of weak beams, it's useful 
to  study it's stability against the listening-in. We don't cover all
possible eavesdropping attacks here, taking into consideration only
the basic listening-in as the preliminary demonstration of the TMCC-channel security 
and protectability.

% pravka 3 (vyshe)

\begin{figure}[htbp]
	\epsfbox{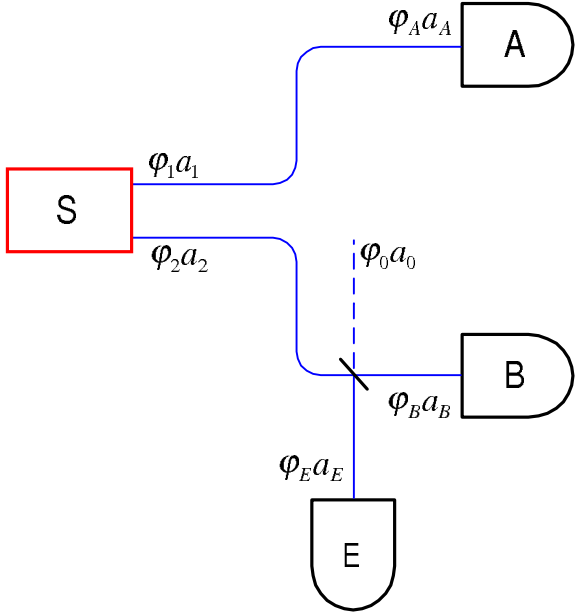}
	\caption{Eavesdropping attack on a TMCC-based quantum channel}
	\label{fig2}
\end{figure}

 Let's assume that some 
eavesdropping intruder (her name is Eve) tries to get the key
being transferred between Alice and Bob through the quantum channel. In 
order to do this, Eve has to split and avert a part of the beam, which goes to 
Bob and detect its intensity by installing a detector at her side (\fref{fig2}). 
The field amplitude of the beam splits then in some $p:q$ ratio and thus 
instead of the quantum mode we have to use the superposition 

\begin{eqnarray}
A = \varphi _A^\ast (x,t)a_A^ + + \varphi _A (x,t)a_A + \varphi _B (x,t)a_B 
+ \varphi _B^\ast (x,t)a_B^ + + \nonumber\\
+ \varphi _E (x,t) a_E  +  \varphi _E^\ast (x,t)a_E^ +
\end{eqnarray}

Obviously, $\varphi _B (x_B ,t_0 ) = 1$,$\varphi _E (x_E ,t_0 ) = 1$ and 
$\varphi _E (x_B ,t_0 ) = 0$, $\varphi _B (x_E ,t_0 ) = 0$, $\varphi _A (x_E 
,t_0 ) = 0$. The 2$^{nd}$ mode is decomposed on the basis, which consists of 
the modes coming to Bob and Eve. In order to describe the properties of 
this beam, we add a mode to the basis of Bob and Eve, which is 
orthogonal to $\varphi _2 $:

\begin{equation}
\varphi _0 = - q\varphi _B + p\varphi _E 
\end{equation}

Without the eavesdropping (and, thus, without the splitter), Eve receives 
only the $\varphi _0 $ mode, in which the laser doesn't radiate, i.e. 
$\varphi _0 = \varphi _E $ and $\varphi _2 = \varphi _B $.

The following conversion of operators corresponds to this decomposition:

\begin{equation}
a_2 = pa_B + qa_E ,
\end{equation}

\begin{equation}
a_0 = - qa_B + pa_E 
\end{equation}

Thus

\begin{equation}
\varphi _0 a_0 + \varphi _2 a_2 = \varphi _B a_B + \varphi _E a_E, 
\end{equation}

\noindent
and similarly for the hermitian-conjugate operators.

These transformations change the state (\ref{eq:tmccground}) to:

\begin{equation}
\left| \lambda \right\rangle = \frac{1}{\sqrt{I_0\left(2\left|\lambda\right|\right)}}I_0 (\lambda a_A^ + (pa_B^ + + qa_E^ + 
)\left| 0 \right\rangle 
\end{equation}

Mean observable values in this case are

\begin{equation}
\left\langle {N_A } \right\rangle = \left\langle \lambda \right|a_A^ + a_A 
\left| \lambda \right\rangle = \frac{\left| \lambda \right|^2I_1 \left( 
{2\left| \lambda \right|} \right)}{I_0 \left( {2\left| \lambda \right|} 
\right)},
\end{equation}

\begin{equation}
\left\langle {N_B } \right\rangle = \left\langle \lambda \right|a_B^ + a_B 
\left| \lambda \right\rangle = p^2\frac{\left| \lambda \right|^2I_1 \left( 
{2\left| \lambda \right|} \right)}{I_0 \left( {2\left| \lambda \right|} 
\right)},
\end{equation}

\begin{equation}
\left\langle {N_E } \right\rangle = \left\langle \lambda \right|a_E^ + a_E 
\left| \lambda \right\rangle = (1 - p^2)\frac{\left| \lambda \right|^2I_1 
\left( {2\left| \lambda \right|} \right)}{I_0 \left( {2\left| \lambda 
\right|} \right)},
\end{equation}

Besides we must take into account the mean values of combinations of 
these operators:

\begin{equation}
\left\langle {N_A N_B } \right\rangle = \left\langle \lambda \right|a_A^ + 
a_A a_B^ + a_B \left| \lambda \right\rangle = p^2\left| \lambda 
\right|^2
\end{equation}

\begin{equation}
\left\langle {N_A N_E } \right\rangle = \left\langle \lambda \right|a_A^ + 
a_A a_E^ + a_E \left| \lambda \right\rangle = (1 - p^2)\left| \lambda 
\right|^2
\end{equation}

\noindent
and

\begin{equation}
\left\langle {N_A ^2} \right\rangle = \left\langle \lambda \right|a_A^ + a_A 
a_A^ + a_A \left| \lambda \right\rangle = \left| \lambda \right|^2
\end{equation}

\begin{equation}
\left\langle {N_B ^2} \right\rangle = \left\langle \lambda \right|a_B^ + a_B 
a_B^ + a_B \left| \lambda \right\rangle = p^4\left| \lambda \right|^2 + 
p^2(1 - p^2)\frac{\left| \lambda \right|^2I_1 \left( {2\left| \lambda \right|} 
\right)}{I_0 \left( {2\left| \lambda \right|} \right)}
\end{equation}

\begin{equation}
\left\langle {N_E ^2} \right\rangle = \left\langle \lambda \right|a_E^ + a_E 
a_E^ + a_E \left| \lambda \right\rangle = (1 - p^2)^2\left| \lambda 
\right|^2 + p^2(1 - p^2)\frac{\left| \lambda \right|^2I_1 \left( {2\left| 
\lambda \right|} \right)}{I_0 \left( {2\left| \lambda \right|} \right)}
\end{equation}

\begin{figure}[htbp]
\begin{center}
	\epsfbox{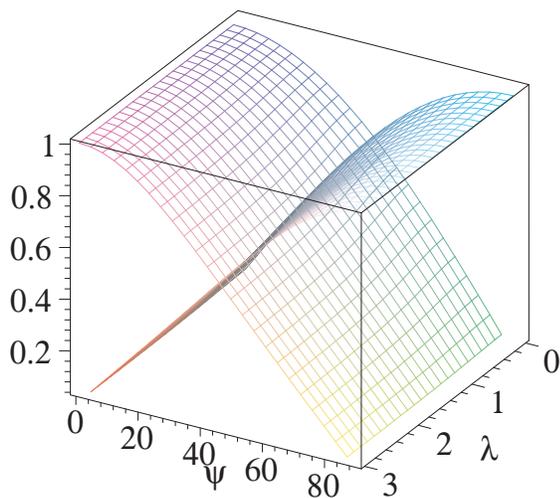}
	\caption{Absolute Alice-Bob and Alice-Eve correlation dependence on the intensity of beam $\lambda$ and the extent of eavesdropping expressed as $p=cos\psi$ for the normalization $p^2+q^2=1$}
	\label{plot1}
\end{center}
\end{figure}
\begin{figure}[htbp]
\begin{center}
	\epsfbox{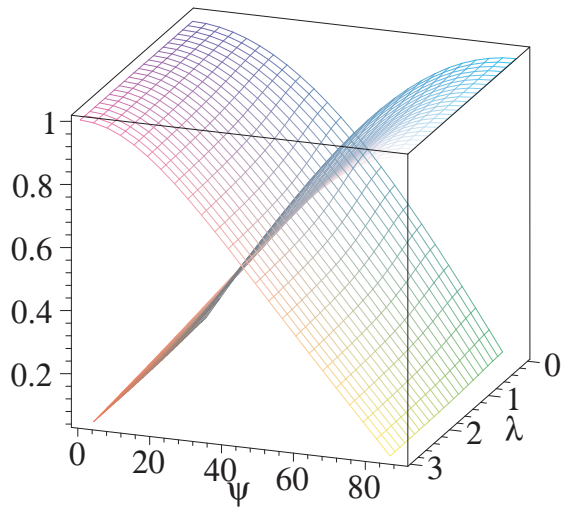}
	\caption{Relative Alice-Bob and Alice-Eve correlation dependence on the intensity   of beam $\lambda$ and the extent $p=cos\psi$ of eavesdropping}
	\label{plot2}
\end{center}
\end{figure}

With these values we can estimate how the Alice-Bob and Alice-Eve correlations depend on the activity of an eavesdropper, which is characterized by 
the parameter $p$ and on the intensity of the beam. The graphs for these 
dependencies are given for both, absolute and relative, correlations in 
\fref{plot1} and \fref{plot2} respectively. One can see that in the case of the weak 
intercept the results of the Bob's measurements almost do not change, but at 
that, if the mean number of photon for Eve in less than 1, she can't really 
distinguish between the 0 bit value and 1, thus the eavesdropping isn't effective. If 
it becomes effective, Bob experiences the same loses in the transmission 
quality, the Alice-Bob correlation becomes significantly less than 1 and the 
channel gets destroyed. This is caused by the fact that each photon, intercepted by
Eve gets absorbed on her detector and thus can't be received by Bob.

%pravka 4 (vyshe)

\section{Conclusions}

Correlated coherent states of the two-mode laser beam (TMCC states) show 
interesting properties, which can be used, in particular, for the tasks of 
the quantum communication and cryptography. 

On the one hand, each of the 
modes looks like a flow of the independent photons rather then a coherent 
beam, since mean values of the operators, which are linear in field, are 
equal to 0 for each mode separately. 

On the other hand, the strong 
correlation between the results of measurements for each of the modes takes 
place. This correlation shows itself in the fact that in each of the modes numbers 
of photons are the same and even the shot noise shows itself equally in the both 
modes. This enables the use of the TMCC state as the generator and carrier 
of random keys. At that, any signficiant attempt of the information intercept in any of 
the channels sharply reduces the correlation, leading to the destruction 
of the channel and, as a consequence, to detection of an eavesdropping. 
Thus, the TMCC-laser generates and transmits exactly the 2 copies of a 
random key. Unlike the single or two-photon schemes, which require large numbers of 
transmission reiterations to obtain the statistically significant 
results, the TMCC beam can be intensive enough to make each single measurement statistically 
significant and thus to use single impulse for each 
piece of information, and remain cryptographically steady. This allows to 
essentially increase the effective data transfer rate and distance.

\end{document}